\documentclass[final,1p,times]{elsarticle}

\usepackage{graphicx,psfrag,color}
\usepackage{dcolumn}
\usepackage{amsmath,amsfonts,amssymb}
\usepackage{bm}
\usepackage{mathbbol}

\def\tr{{\rm tr}\; }

\usepackage{amssymb}
\usepackage{amsthm}
\usepackage{amsmath}

\journal{Annals of Physics}

\begin{document}

\begin{frontmatter}



\title{Optimal continuous variable quantum teleportation protocol for realistic settings}


\author{F. S. Luiz}
\author{Gustavo Rigolin}
\ead{rigolin@ufscar.br}
\address{Departamento de F\'{i}sica,
Universidade Federal de S\~ao Carlos, S\~ao Carlos, SP 13565-905,
Brazil}

\begin{abstract}
We show the optimal setup that allows Alice to teleport coherent states $|\alpha\rangle$ 
to Bob giving the greatest fidelity (efficiency) when one takes into account two realistic
assumptions. The first one is the fact that in
any actual implementation of the continuous variable teleportation protocol (CVTP) Alice and Bob
necessarily share non-maximally entangled states (two-mode finitely squeezed states). 
The second one assumes that Alice's pool of possible coherent states to be teleported to Bob does 
not cover the whole complex plane ($|\alpha|<\infty$). 
The optimal strategy is achieved by tuning three parameters in the original CVTP,
namely, Alice's beam splitter transmittance and Bob's displacements in position and momentum implemented on 
the teleported state. These slight changes in the protocol are currently easy to be implemented and,
as we show, give considerable gain in performance for a variety of possible pool of input states
with Alice.  
\end{abstract}


\end{frontmatter}




\section{Introduction}

The extension of the quantum teleportation protocol from discrete (finite dimensional
Hilbert spaces) \cite{Ben93} to continuous variable (CV) (infinite dimension) 
systems was a landmark to CV quantum communication \cite{Vai94,Bra98,Ral98}.
The main goal of teleportation is to make sure that at the end of the whole 
protocol a quantum state originally describing Alice's system turns out to
describe a quantum system with Bob at a different location.
Moreover, no direct transmission of the system from Alice to Bob is done
and the knowledge of Alice's system
is not needed
at all to accomplish such a task. These two properties clearly 
illustrate why teleportation is so powerful a tool. 
Indeed, for quantum teleportation take place 
Alice and Bob only need to be able to act locally on their systems,
communicate classically and 
share a quantum channel (entangled state). At the end of the process 
Alice's system is no longer described by its original state that
now describes Bob's system.

In principle, a perfect teleportation only occurs   
when Alice and Bob share a maximally entangled state. By perfect teleportation we
mean that at the end of the protocol and with probability one 
Bob's system will be exactly described by the state that originally described
Alice's system.
For discrete systems, and in particular for 
qubits, such maximally entangled states (Bell states) that Alice and Bob must share 
can be experimentally generated in the laboratory
\cite{Bow97,Bos98}.
For CV-systems, however, the perfect implementation of the teleportation protocol 
\cite{Bra98} requires a maximally 
entangled state (the Einstein-Podolsky-Rosen (EPR) state) 
that cannot be generated in the laboratory \cite{Fur98,Bow03a,Kim03}. In modern 
quantum optics terminology, 
one needs an entangled two-mode squeezed state with infinite squeezing ($r \rightarrow \infty$). 
For finite squeezing ($r < \infty$), the teleported quantum state at 
Bob's is never identical to the original
one at Alice's. 

Another assumption in the usual teleportation protocols 
is related to the pool of input states available to Alice, i.e.,
the states that Alice might choose to teleport to Bob. For example, consider the simplest
discrete system, a qubit. In this case it is assumed that Alice's input is given by 
$|\varphi\rangle=a|0\rangle+b|1\rangle$, with $a$ and $b$ random complex numbers 
satisfying the normalization condition $|a|^2+|b|^2=1$. For CV systems, 
and in particular for coherent states $|\alpha\rangle$,
with $\alpha$ complex, it is often assumed that Alice's pool of states cover the entire
complex plane \cite{Bra98,Bra05,Bra00}. From a theoretical point of view,
either for a qubit or a coherent state, these assumptions are the proper ones in order
to determine the strictest conditions guaranteeing a ``truly'' quantum teleportation,
i.e., the conditions where no purely classical protocol can achieve the same 
efficiency as those predicted by the quantum ones \cite{Bra00,Bar98}. From a
practical point of view, however, these assumptions are only valid for qubits, being
unrealistic for CV systems.
Indeed, the energy of a coherent state is proportional to $|\alpha|^2$ and in order 
to cover the entire complex plane we would need states with infinite energy. Also,
the greater $|\alpha|$ the less quantum a coherent state becomes \cite{Bal98}
and other techniques than quantum teleportation such as a direct transmission 
of the state may be better suited in this case.

With these two realistic assumptions in mind, 
a natural question then arises. 
Is it possible to further improve the efficiency of the
standard CV teleportation protocol (CVTP) by taking into account in any modification of the
original setup \cite{Bra98} these two facts? 

For a pool of input coherent states with Alice described by a Gaussian distribution centered at
the vacuum state \cite{Bra01} and when Alice is always teleporting a fixed single state
\cite{Ide02,Mis10,Bow03}, the answer to the previous question is affirmative.
In standard modifications of the CVTP \cite{Bra01,Ide02}, only a single parameter is freely adjusted in
order to improve the quality of the teleported state: Bob is free
to choose the gain $g$ that he might apply equally to the quadratures of his mode
at the end of the protocol (see figure \ref{fig}). 
In the original setup $g=1$, while in the modified versions it was tuned 
as a function of the input states and of the squeezing of the channel in order
to increase the efficiency of  CVTP. An identical strategy was employed to improve the efficiency
of CV entanglement swapping \cite{Pol99,Bra99}, where the optimal $g$ was tuned for a 
specific input state.

What would happen if we go beyond a Gaussian probability distribution centered at the vacuum and
use instead uniform distributions or distributions 
centered in the coherent state $|\beta\rangle$, $\beta \neq 0$?
More important, what are the optimal conditions if we introduce 
\textit{more than one free
parameter} in the modified version of CVTP \cite{Mis10}, 
where either Alice or Bob can change the protocol? 
Our goal here is to investigate these two questions in detail and without assuming we know
the state to be teleported \cite{Mis10,Bow03}. 
The only knowledge we have is the probability of Alice picking a 
particular coherent state $|\alpha\rangle$ according to a predefined 
probability distribution. In other words, Alice and Bob know a priori the
probability distribution describing the possible set of states to be teleported, but
not the actual state at a given execution of the protocol.

In what follows we show that it is possible to achieve further significant increase in performance with 
extra free parameters that introduce, however, minimal changes to the original scheme, 
modifications of which can already be implemented in the laboratory. 
Also, some of the optimal settings are counterintuitive and not found in standard modifications of CVTP. 
For instance, sometimes it is better to use an \textit{unbalanced} beam splitter (BS) instead of 
a balanced one when Alice combines her share of the entangled channel with the state to be teleported.

We also investigate several probability
distributions describing the input states and we show the optimal
modifications to each one of them. Moreover, 
the optimal parameters change appreciably if we work
with either uniform or Gaussian distribution or if those distributions are centered or not on the
vacuum state. And as expected, the changes in the original CVTP not only 
depend on the specific probability distribution
associated to Alice's input states but also on the entanglement of the channel.

\section{Formalism}
\label{secformalism}

\subsection{Qualitative analysis}

Before diving into the mathematical details of our calculations, it is worth 
presenting the bigger picture, i.e., the choices we made from the start in order
to modify the original setup and
the strategy employed to determine the optimal teleportation protocol. 

\begin{figure}[!ht]
\begin{center}
\includegraphics[width=8cm]{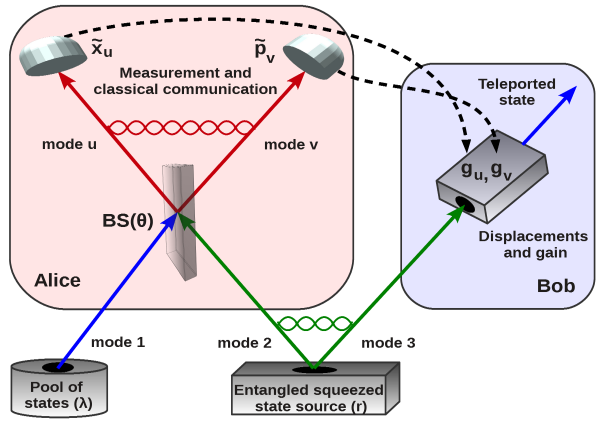}
\caption{\label{fig}  In the original proposal \cite{Bra98} 
we have $\theta=\pi/4$ ($50:50$ BS), 
$g_u=g_v=g\sqrt{2}$, with $g=1$, and position and momentum displacements given by 
$x_3 \rightarrow x_3 + g_u\tilde{x}_u$ and $p_3 \rightarrow p_3 + g_v\tilde{p}_v$. 
These choices yield an average fidelity $F_{av}$ independent of whatever pool of states
(labeled by $\lambda$) is available to Alice. Here, 
for fixed $\lambda$ and squeezing $r$, 
the optimization to get the optimal $F_{av}$ is implemented over 
three free parameters, $\theta, g_v$, and $g_u$, leading to $F_{av}$ 
that depends both on $r$ and $\lambda$. 
See text for details.}
\end{center}
\end{figure}

In the original proposal (see figure~\ref{fig}), 
a two-mode squeezed state with squeezing $r$,
our entanglement resource, is shared between Alice and Bob. Mode $2$ goes
to Alice and mode $3$ to Bob. The state with Alice to be teleported 
is represented by mode $1$, which can be 
any coherent state $|\alpha\rangle$. 
To proceed with the teleportation, Alice 
combines modes $1$ and $2$ in a $50\!\!:\!\!50$ BS and afterwards
measures the position and momentum (quadratures of the electromagnetic field) 
of modes $u$ and $v$, 
respectively, whose results $\tilde{x}_u$ and $\tilde{p}_v$ are then classically
communicated to Bob. With this information he displaces in position 
($x_3 \rightarrow x_3 + g\sqrt{2}\tilde{x}_u$) and 
momentum ($p_3 \rightarrow p_3 + g\sqrt{2}\tilde{p}_v$) 
his mode to get the right teleported state. The displacements and gain 
$(g=1)$ are chosen and fixed as the ones yielding perfect teleportation when
$r \rightarrow \infty$ (maximal entanglement). 

A priory there is no guarantee that the previous choices for the BS transmittance
($\cos^2\theta=1/2$), displacements and gain are the optimal ones for all combinations of 
finite squeezing $r$ and probability distribution for the pool of states available to Alice. 
Therefore, in order to search for the optimal protocol for a given squeezing $r$ and
probability distribution we  
allow the BS to have an arbitrary transmittance ($BS(\theta)$), with 
$0<\theta < \pi/2$ (see figure~\ref{fig}). Furthermore, the quadratures' displacements and gain $g$ 
that Bob must implement 
after Alice informs him of her measurement results ($\tilde{x}_u$ and $\tilde{p}_v$) \cite{Bra98,Bra05}
are also independently chosen in order to optimize the protocol.
Formally, Bob's displacements are given by 
$x_3 \rightarrow x_3 + g_u\tilde{x}_u$ and $p_3 \rightarrow p_3 + g_v\tilde{p}_v$, 
with $g_u$ and $g_v$ chosen in order to optimize the efficiency of CVTP.

\subsection{Quantitative analysis}
\label{quantitative}

\subsubsection{The protocol.}

In what follows we present the details of the mathematical analysis of the modified CVTP, where
the three parameters described above are incorporated into the protocol. We use
interchangeably the words kets, states, and modes to refer to the same object, namely, the
quantized electromagnetic modes \cite{Bra05}. Also, when we refer to 
position $\hat{x}_k=(\hat{a}_k+\hat{a}_k^\dagger)/2$ and momentum 
$\hat{p}_k=(\hat{a}_k-\hat{a}_k^\dagger)/2i$, with $\hat{a}_k$
and $\hat{a}_k^\dagger$ annihilation and creation operators, 
we mean the quadratures of mode $k$,
with commutation relation $[\hat{x}_k,\hat{p}_k]=i/2$.

An arbitrary input state with Alice can be written in the position basis as 
\begin{equation}
|\varphi\rangle= \int dx_1\varphi(x_1)|x_1\rangle,
\label{input}
\end{equation}
where $\varphi(x_1)= \langle x_1 | \varphi \rangle$ and the integration
runs over the entire real line. The entangled state shared between Alice and Bob can 
also be written in the position basis as
\begin{equation}
|\Phi\rangle = \int dx_2 dx_3 \Phi(x_2, x_3) |x_2, x_3 \rangle, 
\label{canal}
\end{equation}
with $\Phi(x_2,x_3)= \langle x_2, x_3 | \Phi \rangle$ and 
$|x_2, x_3 \rangle = |x_2\rangle \otimes |x_3 \rangle$.  Unless stated otherwise,
we keep the ordering of the kets fixed, i.e, the notation $|x_1,x_2,x_3\rangle$ means that
the first two modes/kets are with Alice and the third one with Bob. Using 
equations~(\ref{input}) and (\ref{canal}) we can write the initial state describing all
modes before the beginning of the teleportation protocol as
$|\Psi \rangle = |\varphi\rangle \otimes |\Phi\rangle $, or more explicitly,
\begin{equation}
|\Psi \rangle =\int dx_1 dx_2 dx_3 
\varphi(x_1) \Phi(x_2,x_3)|x_1,x_2,x_3\rangle.
\label{initial}
\end{equation}

The first step in the protocol consists in sending mode 1 (input state) and mode 2 (Alice's
share of the two-mode squeezed state) to a BS with transmittance $\cos^2\theta$
(see figure \ref{fig}). Calling $\hat{B}_{12}(\theta)$ the operator representing the action 
of the BS we have in the position basis \cite{Bra05}
\begin{equation}
\hat{B}_{12}(\theta) |x_1, x_2 \rangle = |x_1 \sin\theta + 
x_2 \cos\theta, x_1 \cos\theta - x_2 \sin\theta\rangle.
\label{bs}
\end{equation}

Inserting equation~(\ref{bs}) into (\ref{initial}) and making the following
variable changes, $x_v = x_1 \sin\theta + x_2 \cos\theta$ and 
$x_u =x_1 \cos\theta - x_2 \sin\theta$, we have
%
%
\begin{eqnarray}
|\Psi' \rangle &=&\int dx_v dx_u dx_3 
\varphi(x_v\sin\theta + x_u\cos\theta)\Phi(x_v\cos\theta - x_u\sin\theta,x_3)|x_v,x_u,x_3\rangle
\label{initial2}
\end{eqnarray}
for the total state after modes $1$ and $2$ go through $BS(\theta)$.

The next step of the protocol consists in measuring the momentum and position 
of modes $v$ and $u$, respectively.  For the quantized electromagnetic mode, this is
achieved by homodyne detectors yielding classical photocurrents that assign real numbers for 
the quadratures $\hat{p}_v$ and $\hat{x}_u$ \cite{Bra05}.

Since Alice will project mode $v$ onto the momentum basis, it is convenient to rewrite
equation~(\ref{initial2}) using the Fourier transformation 
relating the position and momentum eigenstates, 
\begin{equation}
|x_v \rangle = \frac{1}{\sqrt{\pi}} \int dp_v 
e^{- 2i x_v p_v} |p_v\rangle.
\label{fourier1}
\end{equation}

Thus, inserting equation~(\ref{fourier1}) into (\ref{initial2}) we have
%
%
\begin{eqnarray}
|\Psi' \rangle &=&\frac{1}{\sqrt{\pi}}\int dp_v dx_v dx_u dx_3 
\varphi(x_v\sin\theta + x_u\cos\theta)\Phi(x_v\cos\theta - x_u\sin\theta,x_3)e^{- 2i x_v p_v}|p_v,x_u,x_3\rangle.
\nonumber \\
\label{initial2b}
\end{eqnarray}

In the second step of the protocol, Alice measures the momentum of mode $v$ and the position of
mode $u$ (see figure \ref{fig}). Assuming her measurement results are $\tilde{p}_v$ and $\tilde{x}_u$,
the total state at the end of the measurement is simply obtained applying the measurement 
postulate of quantum mechanics, 
$$
|\Psi''\rangle=\hat{P}_{\tilde{p}_v,\tilde{x}_u}|\Psi'\rangle/
\sqrt{\mathbb{p}(\tilde{p}_v,\tilde{x}_u)},
$$
where 
$\hat{P}_{\tilde{p}_v,\tilde{x}_u}=
|\tilde{p}_v,\tilde{x}_u\rangle\langle\tilde{p}_v,\tilde{x}_u| \otimes \mathbb{1}_{3}$
is the projector describing the measurements and 
$\mathbb{1}_3$ is the identity operator acting on mode $3$. 
Also, 
$$
\mathbb{p}(\tilde{p}_v,\tilde{x}_u)=\tr(|\Psi'\rangle\langle\Psi'|\hat{P}_{\tilde{p}_v,\tilde{x}_u})
$$
is the probability of measuring momentum $\tilde{p}_v$ and
position $\tilde{x}_u$, with $\tr$ denoting the total trace. 
Specifying to the position basis and noting that 
$\langle p_v|\tilde{p}_v\rangle=\delta(p_v-\tilde{p}_v)$ and
$\langle x_u|\tilde{x}_u\rangle=\delta(x_u-\tilde{x}_u)$ we have
\begin{equation}
|\Psi'' \rangle = |\tilde{p}_v,\tilde{x}_u\rangle \otimes |\chi'\rangle,
\label{quasebob}
\end{equation}
where Bob's state is 
%
%
\begin{eqnarray}
|\chi' \rangle &=&\frac{1}{\sqrt{\pi \mathbb{p}(\tilde{p}_v,\tilde{x}_u)}}\int dx_v dx_3 
\varphi(x_v\sin\theta + \tilde{x}_u\cos\theta)\Phi(x_v\cos\theta - \tilde{x}_u\sin\theta,x_3)
e^{- 2i x_v \tilde{p}_v}|x_3\rangle.
\label{initial3}
\end{eqnarray}
Here 
\begin{equation}
\mathbb{p}(\tilde{p}_v,\tilde{x}_u) = \int dx_3 |\Psi'(\tilde{p}_v,\tilde{x}_u,x_3)|^2
\label{prob}
\end{equation}
and
%
%
\begin{eqnarray}
\Psi'(\tilde{p}_v,\tilde{x}_u,x_3)=\langle \tilde{p}_v,\tilde{x}_u,x_3| \Psi'\rangle
=\frac{1}{\sqrt{\pi}}\int dx_v
\varphi(x_v\sin\theta + \tilde{x}_u\cos\theta)\Phi(x_v\cos\theta - \tilde{x}_u\sin\theta,x_3)e^{- 2i x_v \tilde{p}_v},
\label{psilinha}
\end{eqnarray}
where equation~(\ref{psilinha}) was obtained using (\ref{initial2b}).

The third step of the protocol consists in Alice sending to Bob via a classical channel 
(photocurrents) her measurement results. With this information Bob is able to implement the fourth
and last step of the protocol, namely, he displaces his mode quadratures according to the following
rule, $x_3 \rightarrow x_3 + g_u\tilde{x}_u$ and $p_3 \rightarrow p_3 + g_v\tilde{p}_v$. 
Mathematically, this corresponds to the application of the displacement operator 
$\hat{D}(\alpha)=e^{\alpha \hat{a}^\dagger-\alpha^*\hat{a}}=
e^{-2iRe[\alpha]\hat{p}+2iIm[\alpha]\hat{x}}$, with 
$\alpha = g_u\tilde{x}_u + ig_v\tilde{p}_v$. Since $\hat{x}$ and $\hat{p}$ commute with
their commutator we can apply Glauber's formula to obtain 
$\hat{D}(\alpha)=e^{iRe[\alpha]Im[\alpha]}e^{-2iRe[\alpha]\hat{p}}e^{2iIm[\alpha]\hat{x}}$.
This leads to
\begin{equation}
\hat{D}(g_u\tilde{x}_u + ig_v\tilde{p}_v)|x_3\rangle = e^{ig_ug_v\tilde{x}_u\tilde{p}_v}e^{2ig_v\tilde{p}_vx_3}
|x_3 + g_u\tilde{x}_u\rangle.
\label{disp}
\end{equation}

The final state with Bob,
$|\chi\rangle=\hat{D}(g_u\tilde{x}_u + ig_v\tilde{p}_v)|\chi'\rangle$, can be put as follows if
we use equation~(\ref{disp}) and make the variable change
$x_3 \rightarrow x_3 - g_u\tilde{x}_u$,
%
\begin{eqnarray}
|\chi \rangle &=&\frac{e^{-ig_ug_v\tilde{x}_u\tilde{p}_v}}{\sqrt{\pi \mathbb{p}(\tilde{p}_v,\tilde{x}_u)}}\int dx_v dx_3 
\varphi(x_v\sin\theta + \tilde{x}_u\cos\theta)\nonumber \\
&&\times \Phi(x_v\cos\theta - \tilde{x}_u\sin\theta,x_3-g_u\tilde{x}_u)
e^{- 2i (x_v -g_vx_3)\tilde{p}_v}|x_3\rangle\nonumber \\
&=&\int dx_3 \left(\frac{e^{-ig_ug_v\tilde{x}_u\tilde{p}_v}}{\sqrt{\pi \mathbb{p}(\tilde{p}_v,\tilde{x}_u)}}
\int dx_v\varphi(x_v\sin\theta + \tilde{x}_u\cos\theta)\right.\nonumber \\
&&\left.\times \Phi(x_v\cos\theta - \tilde{x}_u\sin\theta,x_3-g_u\tilde{x}_u)
e^{- 2i (x_v -g_vx_3)\tilde{p}_v}\right)|x_3\rangle\nonumber \\
&=& \int dx_3 \chi(x_3) |x_3\rangle.
\label{bobfinal}
\end{eqnarray}
Note that $e^{-ig_ug_v\tilde{x}_u\tilde{p}_v}$ in Bob's final state is an 
irrelevant global phase and could be suppressed.

It is worth mentioning at this point that equation~(\ref{bobfinal}), together with 
equations~(\ref{prob}) and (\ref{psilinha}), are quite general. They allow us to get the teleported
state with Bob for any input state and any entangled state (channel) shared 
between Alice and Bob. For instance, if we use 
a maximally entangled state (EPR state) we have \cite{Bra05} 
$\Phi(x_2,x_3)\propto \delta(x_2-x_3)$ in equation~(\ref{canal}). Using this channel
in equation~(\ref{bobfinal}) leads to $\chi(x_3) = \varphi(x_3)$ if we set
$g_u=g_v=\sqrt{2}$ and $\theta=\pi/4$, i.e, 
we have a perfect teleportation.

\subsubsection{Fidelity.}
\label{nslbfid}

As a figure of merit to decide the optimality of the protocol 
we employ the average fidelity $F_{av}$ \cite{Bra00},
whose computation requires the specification of the presumed probability 
of available states with Alice, assumed fixed throughout the many runs of the protocol. 
The fidelity measures how close the output state with Bob at the end of the protocol
is to the input state employed by Alice. Here the average is taken over the fidelities of 
each input state and its respective output, with the weight of each state given
by its probability to be picked out of the states available to Alice.  
Our strategy consists, therefore, in choosing $\theta, g_v$, and $g_u$ in order
to maximize $F_{av}$.  

Mathematically, if the density matrix describing the output state with Bob after one single run of the protocol
is $\hat{\rho}_B=|\chi \rangle\langle\chi|$, the fidelity is defined as
\begin{equation}
F(|\varphi\rangle,\tilde{p}_v,\tilde{x}_u)=\langle \varphi | \hat{\rho}_B | \varphi \rangle,
\label{fid1}
\end{equation}
where we highlight that the fidelity depends on the input state $|\varphi\rangle$
and on the measurement outcomes obtained by Alice. For the moment we leave implicit the
dependence on the other parameters, i.e, 
what channel/entanglement we have and $\theta$, $g_v$, and $g_u$. 
Note that $F$  achieves its maximal value (one) only if we
have a flawless teleportation ($\hat{\rho}_B = \hat{\rho}_{input}$) and its minimal one (zero)
if the output is orthogonal to the input.

At each run of the protocol, Alice will measure  $\tilde{p}_v$ and $\tilde{x}_u$ with probability
$\mathbb{p}(\tilde{p}_v,\tilde{x}_u)$. Hence, Bob's final ensemble of states, 
averaged over all possible measurement results for a fixed input state
$|\varphi\rangle$, is
\begin{equation}
F(|\varphi\rangle)=\int d\tilde{p}_vd\tilde{x}_u \mathbb{p}(\tilde{p}_v,\tilde{x}_u)
F(|\varphi\rangle,\tilde{p}_v,\tilde{x}_u).
\label{fid2}
\end{equation}
 
Furthermore, to properly search the optimal configuration for a probability
distribution of input states $P(|\varphi\rangle)$ with Alice, another averaging
is needed, this time over the pool of states available to her \cite{Bra00},
\begin{equation}
F_{av}(\theta,g_v,g_u)=\int d|\varphi\rangle P(|\varphi\rangle) F(|\varphi\rangle).
\label{fid3}
\end{equation}

The strategy to optimize the teleportation protocol, once we know $P(|\varphi\rangle)$,
is the search for the triple of points $(\theta,g_v,g_u)$ maximizing $F_{av}(\theta,g_v,g_u)$. 
Therefore, for fixed entanglement, we either solve analytically (if possible) or numerically 
the following three equations for $\theta,g_v,$ and $g_u$,
\begin{equation}
\frac{\partial F_{av}}{\partial \theta} = 0, \hspace{.3cm} \frac{\partial F_{av}}{\partial g_v} = 0,
\hspace{.3cm}\frac{\partial F_{av}}{\partial g_u} = 0.
\label{max}
\end{equation}
The obtained solutions are then inserted into (\ref{fid3}) and slightly varied about their actual
values in order to be sure we have a global maximum, ruling out possible 
local maximums, minimums, and saddle points.

\section{Results}
\label{secresults}

In the rest of this paper we particularize to 
input states at Alice's given by coherent states,
$$
|\varphi\rangle=|\alpha\rangle,
$$ 
with $\alpha$ complex, and to the entanglement shared between
Alice and Bob given by two-mode squeezed vacuum states, 
$$
|\Phi\rangle = |\psi_r\rangle = 
\sqrt{1-\tanh^2r}\sum_{n=0}^{\infty}\tanh^nr|n\rangle_A\otimes|n\rangle_B,
$$
where $|n\rangle_{A(B)}$ are Fock number states at Alice's (Bob's). Also, $r$ is the squeezing
parameter and for $r=0$ we have $|00\rangle$, the vacuum state, and for $r\rightarrow \infty$ the
unphysical maximally entangled EPR state. 
Both coherent and two-mode squeezed states of the quantized electromagnetic
field are easily generated in the laboratory today and they were the ingredients employed
in the experimental implementation of the original CVTP \cite{Fur98}. 

These two states are represented in the position basis as \cite{Bra05}
\begin{equation}
\phi(x_1)=\langle x_1 | \alpha \rangle = \left(\frac{2}{\pi}\right)^{1/4}e^{-x_1^2+2\alpha x_1
-|\alpha|^2/2-\alpha^2/2}
\label{in}
\end{equation}
and
\begin{eqnarray}
\Phi(x_2,x_3)&=& \langle x_2,x_3 | \psi_r \rangle =
\sqrt{\frac{2}{\pi}}\exp\left[-e^{-2r}(x_2+x_3)^2/2 - e^{2r}(x_2-x_3)^2/2\right].
\label{ch}
\end{eqnarray}

Equations (\ref{in}) and (\ref{ch}) allow us to explicitly compute equation~(\ref{fid2}),
%
%
\begin{eqnarray}
F(|\alpha\rangle) = f_2(\theta,g_v)^{-\frac{1}{2}}
\exp\left[-\frac{f_1(\theta,g_v)}{f_2(\theta,g_v)}Im[\alpha]^2\right]
f_2\hspace{-.1cm}
\left(\theta-\frac{\pi}{2},g_u\right)^{\hspace{-.1cm}-\frac{1}{2}}
\exp\left[-\frac{f_1(\theta+\frac{\pi}{2},g_u)}
{f_2(\theta-\frac{\pi}{2},g_u)}Re[\alpha]^2
\right],
\label{fid2b}
\end{eqnarray}
where
%
%
\begin{eqnarray}
f_1(\theta,g_v)&=&(1 - g_{v} \sin\theta)^2, \\
f_2(\theta,g_v)&=& [(2 + g_v^2) \cosh^2r + g_v^2\cos(2\theta)\sinh^2r - 2g_v \cos\theta\sinh(2r)]/2.
\end{eqnarray}

Looking at equation~(\ref{fid2b}) we see that if we want to have $F(|\alpha\rangle)$ independent of
$|\alpha\rangle$ we have to choose $g_v$ and $g_u$ such that 
$f_1(\theta,g_v)=f_1(\theta+\pi/2,g_u)=0$. This is accomplished if $g_v=\csc\theta$ and 
$g_u=\sec\theta$. Inserting these values for $g_v$ and $g_u$ into equation~(\ref{fid2b}) and maximizing
it we get $\theta=\pi/4$ and $g_v=g_u=\sqrt{2}$, as the optimal 
parameter configuration, and $F(|\alpha\rangle)=1/(1+e^{-2r})$ for the optimal fidelity.
These are the configuration and fidelity
of the original CVTP. However, when we maximize the fidelity taking into account a specific 
$|\alpha\rangle$, or a pool of states $|\alpha\rangle$ with Alice,
the optimal setting necessarily changes. 

In what follows we work with several different probability distributions
$P(|\alpha\rangle)$ for Alice's pool of states, whose normalization condition reads 
\begin{eqnarray}
\int P(|\alpha\rangle)d|\alpha\rangle&=&\int_{-\infty}^{\infty}\int_{-\infty}^{\infty} 
P(\alpha)dRe[\alpha] dIm[\alpha]
\nonumber \\
&=&\int_{0}^{2\pi}\int_{0}^{\infty} P(\alpha)|\alpha|d|\alpha| d\omega=1,
\end{eqnarray}
where $\alpha= Re[\alpha]+iIm[\alpha]=|\alpha| e^{i\omega}$.

\subsection{Purely real or imaginary states}

The first two distributions we work with confine the states $|\alpha\rangle$
to be given by either real or imaginary $\alpha$. We assume the states to
be uniformly distributed along the real or imaginary axis from
$-R$ to $R$, where $R>0$. Real and imaginary states are employed in CV quantum cryptography,
where the encoding of the key is given by those states \cite{Ral99,Gra02,Leu02,Gra03,Nam03,Nam03b,Nam04,Nam06,Leu09,Leu10,Jou13}.

The real $\alpha$ distribution is
given by 
\begin{equation}
P_r(\alpha)=\delta(Im[\alpha])\Theta(R^2 - Re[\alpha]^2)/2R,
\label{pr}
\end{equation} 
with $\delta(x)$ being the Dirac delta function and 
$\Theta(x)$ the Heaviside theta function ($\Theta(x)=0$ if $x<0$ and 
$\Theta(x)=1$ for $x\geq0$). The imaginary $\alpha$ distribution
reads  
\begin{equation}
P_i(\alpha)=\delta(Re[\alpha])\Theta(R^2 - Im[\alpha]^2)/2R.
\label{pi}
\end{equation}

Inserting equation~(\ref{pr}) into (\ref{fid3}) we obtain for the average 
fidelity of states on the real line,
\begin{equation}
F_{av}^r(\theta,g_v,g_u)=\frac{\sqrt{\pi}}{2R} \frac{\mbox{Erf}
\left[R\sqrt{\frac{f_1(\theta+\pi/2,g_u)}
{f_2(\theta-\pi/2,g_u)}}\right]}{[f_1(\theta+\pi/2,g_u)f_2(\theta,g_v)]^{1/2}},
\label{favr}
\end{equation}
where $\mbox{Erf}[x]=\frac{2}{\sqrt{\pi}}\int_{0}^{x}e^{-t^2}dt$ 
is the error function. 

Since the dependence of (\ref{favr}) on $g_v$ is simply given by 
$f_2(\theta,g_v)^{-1/2}$, it is straightforward to solve for 
$\partial{F_{av}^r/\partial{g_v}}=0$, leading to the following
optimal $g_v$,
\begin{equation}
g^{opt}_v = \frac{\sinh (2r) \cos \theta^{opt}}{\cosh^2 r + \cos (2\theta^{opt})\sinh^2 r }.
\label{gvropt}
\end{equation}

Now, substituting equation~(\ref{gvropt}) into (\ref{favr}) we obtain the optimal average fidelity
by solving the remaining two equations in (\ref{max}). However, due to the presence
of the error function in (\ref{favr}), we cannot get a closed solution for $g_u^{opt}$ and
$\theta^{opt}$. Thus, we must rely on numerical solutions once the squeezing $r$ 
and  the range $R$ of the distribution are specified.

Similarly, for imaginary states we have
\begin{equation}
F_{av}^i(\theta,g_v,g_u)=\frac{\sqrt{\pi}}{2R} \frac{\mbox{Erf}
\left[R\sqrt{\frac{f_1(\theta,g_v)}
{f_2(\theta,g_v)}}\right]}{[f_1(\theta,g_v)f_2(\theta-\pi/2,g_u)]^{1/2}}.
\label{favi}
\end{equation}

Note that the roles of $g_v$ and $g_u$ are now interchanged and the extremum condition 
$\partial{F_{av}^i/\partial{g_u}}=0$ is easily solved and gives the following 
optimal $g_u$,
\begin{equation}
g^{opt}_u = \frac{\sinh (2r) \sin \theta^{opt}}{\cosh^2 r - \cos (2\theta^{opt})\sinh^2 r }.
\label{guiopt}
\end{equation}
The remaining two extremum conditions must be numerically computed.

\begin{figure}[!ht]
\begin{center}
\includegraphics[width=8cm]{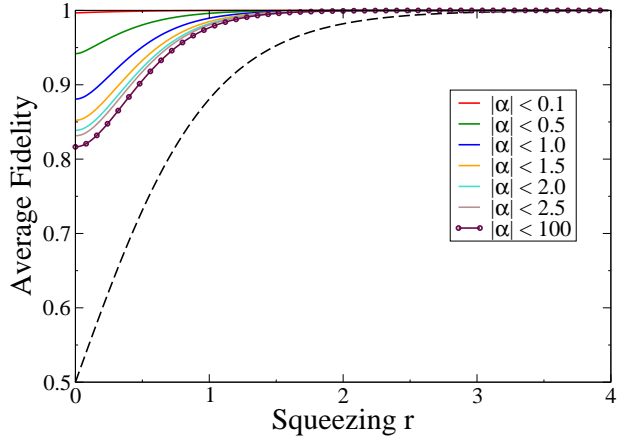}
\caption{\label{fig4}  Solid curves give the optimal average fidelities
as functions of the entanglement of the channel (squeezing $r$) 
for real and imaginary uniform distributions, whose ranges $R=|\alpha|$ increase from
top to bottom. Dashed curve: average fidelity given
by the original CVTP, which does not depend on a particular distribution.
Note that for distributions covering the entire real or imaginary line
($R\rightarrow \infty$, maroon/circle curve) we still get impressive gains in 
efficiency when dealing with channels possessing a low degree of entanglement.}
\end{center}
\end{figure}
\begin{figure}[!ht]
\begin{center}
\includegraphics[width=8cm]{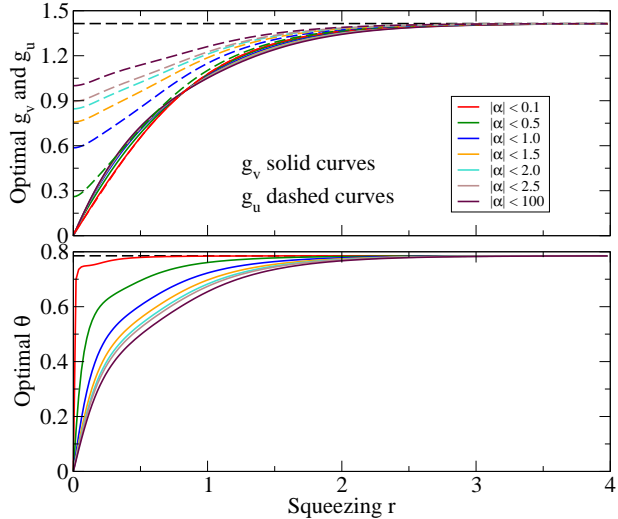}
\caption{\label{fig5}  Parameters giving the
optimal average fidelities shown in figure \ref{fig4} for states lying on the
real line. Note that the many curves for $g_v^{opt}$ are very close to each other.
For $g_u^{opt}$ the range $R=|\alpha|$ increases from bottom to top while for 
$\theta$ it increases from top to bottom.
The dashed black curves give the values used in the original CVTP
($g_v=g_u=\sqrt{2}\approx 1.41$ and $\theta=\pi/4\approx 0.79$). 
For the imaginary distribution, $g_v\leftrightarrow g_u$ and 
$\theta \rightarrow \pi/2 - \theta$ in the graphics above.}
\end{center}
\end{figure}

In figure \ref{fig4} we show the optimal $F_{av}^r(\theta,g_v,g_u)$
and $F_{av}^i(\theta,g_v,g_u)$ for several distributions with $|\alpha|\leq R$
as a function of the squeezing $r$ of the channel. It is worth mentioning that for
$r=0$ we have no entanglement shared between Alice and Bob and the optimal values for
the fidelity are the ones one would get by using only classical resources.

The first thing we note looking at figure \ref{fig4} is that the optimal average fidelities are the same
for the real and imaginary distributions. Second, the lower the range $R$ of
the distribution the greater the efficiency. 
This is expected since as we decrease $R$ the states available
to Alice become more and more similar. Therefore, the optimal parameters giving
the best average fidelity approach the optimal parameters giving the highest fidelity
for each one of the states within the distribution. Third, as we increase $R$ the
optimal fidelity decreases. However, it rapidly tends to its asymptotic limit
(maroon/circle line in figure \ref{fig4}), which is still 
by far superior than the 
fidelity given by the original CVTP (dashed line in figure \ref{fig4}).
Figure \ref{fig5} gives the optimal $\theta, g_v$, and $g_u$ leading to the
optimal average fidelities shown in figure \ref{fig4}. Note that for low
squeezing, the optimal $\theta$ is not $\pi/4$. This means that Alice needs to
use an unbalanced BS to get the optimal fidelity.

\begin{figure}[!ht]
\begin{center}
\includegraphics[width=8cm]{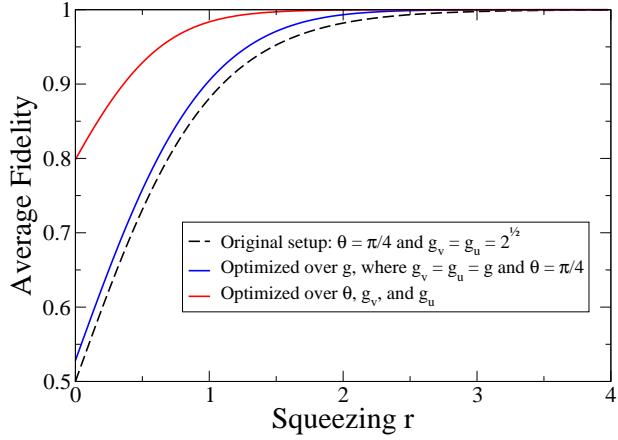}
\caption{\label{fig6}  The curves above were calculated considering
either a real or imaginary uniform distribution with $|\alpha|\leq R = 5.0$. It is clear
from the figure that to get an expressive
gain in efficiency 
for quantum channels with low degree of entanglement it is crucial to optimize
the average fidelity over the three free parameters.}
\end{center}
\end{figure}

We have also compared how much we gain in efficiency optimizing
$F_{av}(\theta,g_v,g_u)$ considering $\theta,g_v$, and $g_u$ as
free parameters against  the optimization of $F_{av}(\pi/4,g,g)$, 
where only the gain $g$ is free \cite{Bra01,Ide02,Bra99}. 
As can be seen in figure \ref{fig6},
we obtain a considerable gain in efficiency when we allow the three
parameters to be freely adjusted for a given distribution when compared
to the single parameter scenario. Also, we have checked that the 
greater $R$ the less efficient is the single parameter optimization. For not 
too big $R$ it approaches the original CVTP fidelity 
and we must resort to 
the three parameter optimization to get effective efficiency gains.

\subsection{States lying on a circumference or a disk}

For a set of states $|\alpha\rangle$ 
available to Alice having the same
amplitude $|\alpha|=R$ and phases $\omega$ given by a uniform distribution, where
$0\leq\omega < 2\pi$, we have in the complex plane representation of $|\alpha\rangle$
a circumference of radius $R$ centered in the vacuum. Such states with fixed magnitude 
and phases randomly chosen 
are used in the encoding of random keys in quantum cryptography based on CV states \cite{Ral99,Gra02,Leu02,Gra03,Nam03,Nam03b,Nam04,Nam06,Leu09,Leu10,Jou13}. 
The determination of the optimal settings for the teleportation of such states is important since one can use the CVTP instead of
the direct sending of those states from Alice to Bob in order to generate a secrete key.

The probability distribution representing such states can be written as
\begin{equation}
P_c(\alpha) = \delta(|\alpha|-R)/(2\pi R)
\label{pc}
\end{equation}
and inserting equation~(\ref{pc}) into (\ref{fid3}) leads to
\begin{eqnarray}
F_{av}^{c}(\theta,g_v,g_u) =
\frac{e^{-h_+(\theta,g_v,g_u)R^2}I_0[h_-(\theta,g_v,g_u)R^2]}
{\sqrt{f_2(\theta,g_v) f_2(\theta-\pi/2,g_u)}}.
\label{favc}
\end{eqnarray}
Here
\begin{eqnarray}
h_\pm(\theta,g_v,g_u) =  \left(\frac{f_1(\theta+\pi/2,g_u)}
{2f_2(\theta-\pi/2,g_u)}\pm\frac{f_1(\theta,g_v)}{2f_2(\theta,g_v)}\right)
\end{eqnarray}
and $I_0[x]$ is the modified Bessel function of the first kind, i.e, 
the solution to $x^2I_n''[x]+xI_n'[x]-(x^2+n^2)I_n[x]=0$ with $n=0$ and
boundary conditions $I_0[0]=1$ and 
$\lim_{x\rightarrow \pm\infty}I_0[x]\rightarrow \infty$. 

The existence of the Bessel function makes the analytic solution
of the optimization problem unfeasible. However, we have numerically
checked for many random combinations of $r$ and $R$ 
that all set of parameters $\theta, g_v$, and $g_u$ leading to the
optimal average fidelity for this distribution are such that
$\theta=\pi/4$ and $g_v=g_u=g$. Therefore, inserting the previous parameters
into equation~(\ref{favc}), we can recast the determination of the optimal 
average fidelity to a single variable optimization problem. 
A direct computation gives the following optimal
average fidelity,
\begin{eqnarray}
\mathcal{F}_{opt}^{c}(g) =
\frac{2 \exp \left[-\frac{\left(\sqrt{2} g-2\right)^2 R^2 \mbox{\scriptsize{sech}}\; r}
{2 \left(g^2+2\right) \cosh r-4 \sqrt{2} g \sinh r}\right]}
{( g^2 + 2 ) \cosh ^2 r-\sqrt{2} g \sinh (2 r)},
\label{favcopt}
\end{eqnarray}
where the optimal $g$ is given by solving the
cubic equation
\begin{eqnarray*}
\sqrt{2} (e^r\! \sinh (2 r\!) \cosh r+2 R^2)-g e^r\! (3 \cosh (2 r\!)-1) \!\cosh r
\\
-g^2 \sqrt{2}(R^2-3 e^r \sinh r \cosh ^2r)-g^3 e^r \cosh ^3r =0.
\end{eqnarray*}

\begin{figure}[!ht]
\begin{center}
\includegraphics[width=8cm]{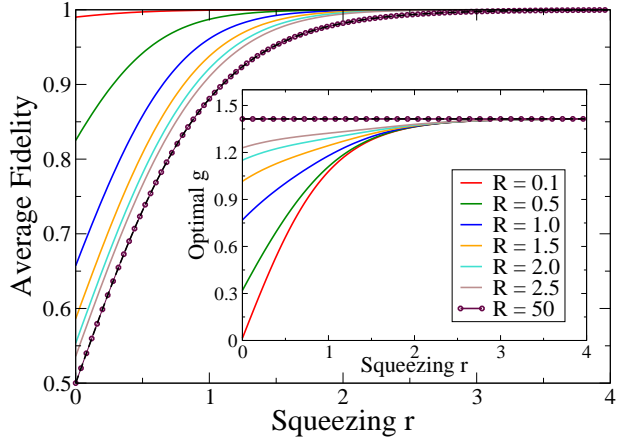}
\caption{\label{fig9}  Solid curves are the optimal average fidelities
as functions of the entanglement of the channel (squeezing $r$) for states uniformly
lying on a circumference of radius $R$, with
$R$ increasing from top to bottom. 
Inset: Optimal gain $g_v=g_u=g$ 
giving the optimal fidelities shown in the main
graph. Here, $R$ increases from bottom to top.
Dashed curves: average fidelity (main graph) and $g$ (inset)
according to the original CVTP. The dashed
curves are indistinguishable from the $R=50$ curves.
}
\end{center}
\end{figure}

In figure \ref{fig9} we plot the optimal $F_{av}^c(\theta,g_v,g_u)$,
or equivalently $\mathcal{F}_{opt}^{c}(g)$,
for several values of $R$ as a function of the entanglement of the channel. 

Looking at figure \ref{fig9} we see that 
the smaller the radius $R$ of the distribution the greater the efficiency. Also,
for $R=0$ we have $\mathcal{F}_{opt}^{c}(g)=1$ for any value of squeezing
since in this case we just have one state to teleport, the vacuum state. 
In this scenario Bob has complete knowledge of what state will be sent and the 
teleportation is trivial. 
Now, as we increase $R$ the optimal fidelity decreases, and contrary
to the pure real and imaginary distributions, it does not tend to an asymptotic limit
(maroon/circle line in figure \ref{fig9}) yielding a better performance than that
given by the original CVTP. Actually, $\mathcal{F}_{opt}^{c}(g)$ approaches 
the fidelity of the original CVTP. 

We have also computed the optimal average fidelities assuming 
that both the amplitude and phase are given
by independent uniform distributions, i.e., the input states are 
contained in a disk of radius 
$R$ ($|\alpha|\leq R$ and $0\leq\omega < 2\pi$). In this case
$
P_d(\alpha) = \Theta(R-|\alpha|)/(\pi R^2).
$
After a systematic numerical study for several values
of squeezing $r$ and discs with radius $R$ we obtained that the optimal parameters
are such that $\theta^{opt}=\pi/4$ and $g_v^{opt}=g_u^{opt}=g$, leading to the optimal average fidelity 
\begin{equation}
\mathcal{F}_{opt}^{d}(g) =
\frac{2 \left(1-\exp \left[\frac{-\left(\sqrt{2}-g\right)^2 R^2}
{(2+g^2) \cosh ^2(r)-\sqrt{2} g \sinh (2 r)}\right]\right)}
{\left(\sqrt{2}-g\right)^2 R^2}.
\label{favdopt}
\end{equation}
The optimal fidelity as well as the optimal settings possess the
same qualitative features already explained for the distribution of states on a circumference. 
Quantitatively, however, we have a better performance for a given disc of radius $R$ 
when compared to a circumference of the same radius. This is understood noting that a disc
is the union of all circumferences with radius lower than or equal
to $R$. And since we have shown that the smaller $R$ the greater the 
fidelity for a circumference distribution,  it is clear that the optimal average fidelity of a disc 
should outperform the optimal one for a circumference with the same $R$.

\subsection{States given by Gaussian distributions}

We now consider that the pool of input states with Alice is described by a 
Gaussian distribution with variance $1/(2\lambda)$ and mean $\beta$, 
\begin{equation}
P_g(\alpha)=(\lambda/\pi)\exp{(-\lambda|\alpha-\beta|^2)}.
\label{pg}
\end{equation}
Here, when $\beta=0$ the distribution is centered at the vacuum state and
for $\beta\neq 0$ it is centered at the coherent state $|\beta\rangle$.
Also, as we increase $\lambda$ (decrease the variance) the
distribution approaches a single point, $\beta$, and for $\lambda \rightarrow 0$ we
have a uniform distribution covering the entire complex plane.

These distributions represent what one actually gets when trying to generate a given coherent state $|\beta\rangle$.
Indeed, one is never able to exactly generate a coherent state with the exact $\beta$. Rather, the generated state lies within 
a Gaussian distribution about the desired state, whose width depends on the quality of the coherent 
state generation scheme.

Inserting equation~(\ref{pg}) into (\ref{fid3}) we can readily compute the average fidelity for 
an input of coherent states distributed according to a Gaussian centered at $\beta$,
%
%
\begin{eqnarray}
F_{av}^{g}(\theta,g_v,g_u) =
\frac{\sqrt{\lambda}\exp\left[\frac{-\lambda f_1(\theta + \pi/2,g_u) Re[\beta]^2}
{f_1(\theta + \pi/2,g_u)+\lambda f_2(\theta - \pi/2,g_u)}\right]
\sqrt{\lambda}  
\exp\left[\frac{-\lambda f_1(\theta,g_v) Im[\beta]^2 }{f_1(\theta,g_v)+\lambda f_2(\theta,g_v) }\right]
}
{
\sqrt{f_1(\theta + \pi/2,g_u)+\lambda f_2(\theta - \pi/2,g_u)}
\sqrt{f_1(\theta,g_v)+\lambda f_2(\theta,g_v) }
}.
\label{favg}
\end{eqnarray}

For Gaussians with $\beta=0$ it is not difficult to show that the optimal average fidelity is such
that $\theta^{opt}=\pi/4$ and $g_v=g_v=g$, where $g$ is \cite{Bra01} 
\begin{equation}
g = \frac{2\sqrt{2}+\lambda\sqrt{2}\sinh (2r)}{2+\lambda+\lambda\cosh(2r)}.
\label{optGgauss}
\end{equation}
Note that as the Gaussian distribution variance increases, approaching a uniform distribution 
covering the whole complex plane ($\lambda \rightarrow 0$), equation (\ref{optGgauss}) gives
$$
\lim_{\lambda \rightarrow 0}g = \sqrt{2}.
$$
This is exactly the value of $g$ for the original
CVTP and illustrates that it is a particular case of the generalized CVTP here
presented.

\begin{figure}[!ht]
\begin{center}
\includegraphics[width=8cm]{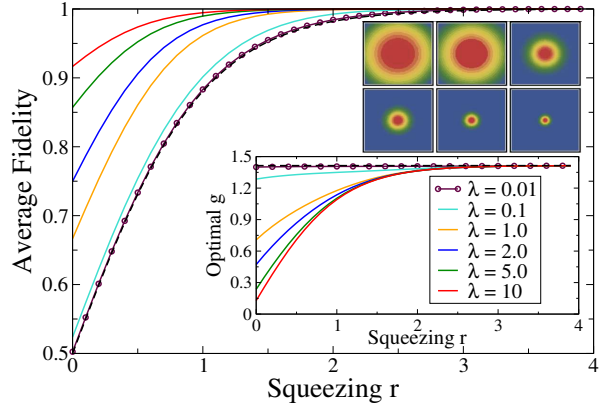}
\caption{\label{fig18}  Solid curves give the optimal average fidelities
as functions of the entanglement of the channel (squeezing $r$) for a pool of input 
states given by Gaussians centered at the origin with variance $1/(2\lambda)$, with $\lambda$ increasing from 
bottom to top. Dashed curve: average fidelity given
by the original CVTP, which is indistinguishable from the corresponding one for
the Gaussian with $\lambda = 0.01$.
Upper inset: Density plots of the several Gaussian distributions. 
Their variance ($\lambda$) decreases (increases) from left to right. 
Lower inset: The optimal gain $g_v=g_u=g$ 
giving the optimal fidelities shown in the main
graph. Here, $\lambda$ increases from top to bottom and the dashed
curve is $g$ according to the original CVTP, indistinguishable from the optimal one 
for the Gaussian with $\lambda = 0.01$. The optimal $\theta$ is always $\pi/4$.}
\end{center}
\end{figure}

In figure \ref{fig18} we show how the optimal average fidelity depends on the entanglement of the 
channel for several Gaussians with different variances. As expected, as we decrease $\lambda$,
covering the entire complex plane, we recover the results of the original CVTP.

\begin{figure}[!ht]
\begin{center}
\includegraphics[width=8cm]{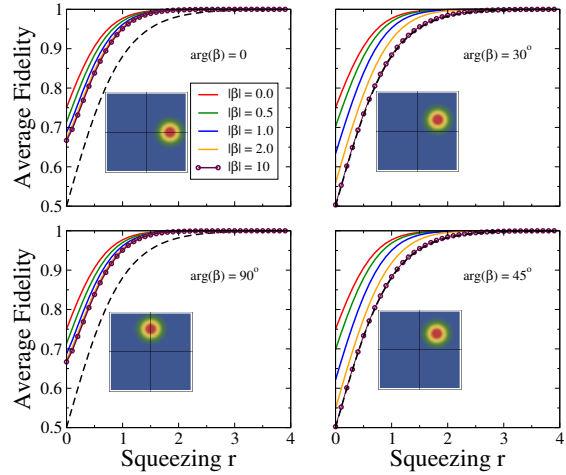}
\caption{\label{fig19}  Optimal average
fidelity as a function of the squeezing $r$ of the channel
for Gaussian distributions with variance
$1/(2\lambda)$, $\lambda = 2.0$, and mean $\beta = |\beta|e^{\arg(\beta)}$. 
$|\beta|$ increases from top to bottom (solid curves) and $\arg(\beta)$ are shown in the
graphics and illustrated in the insets. The dashed curve gives the fidelity of the original
CVTP.
For low squeezing, note that as we increase $|\beta|$ for $\arg(\beta)=0$ or $\pi/2$ (left panels)
the optimal average fidelity tends to values far superior than that predicted by the original
CVTP. This interesting fact does not happen if the center of the distribution moves away from
the real or imaginary axis (right panels), where the curves for the original
CVTP and $|\beta| = 10$ fidelities cannot be distinguished.
}
\end{center}
\end{figure}
\begin{figure}[!ht]
\begin{center}
\includegraphics[width=8cm]{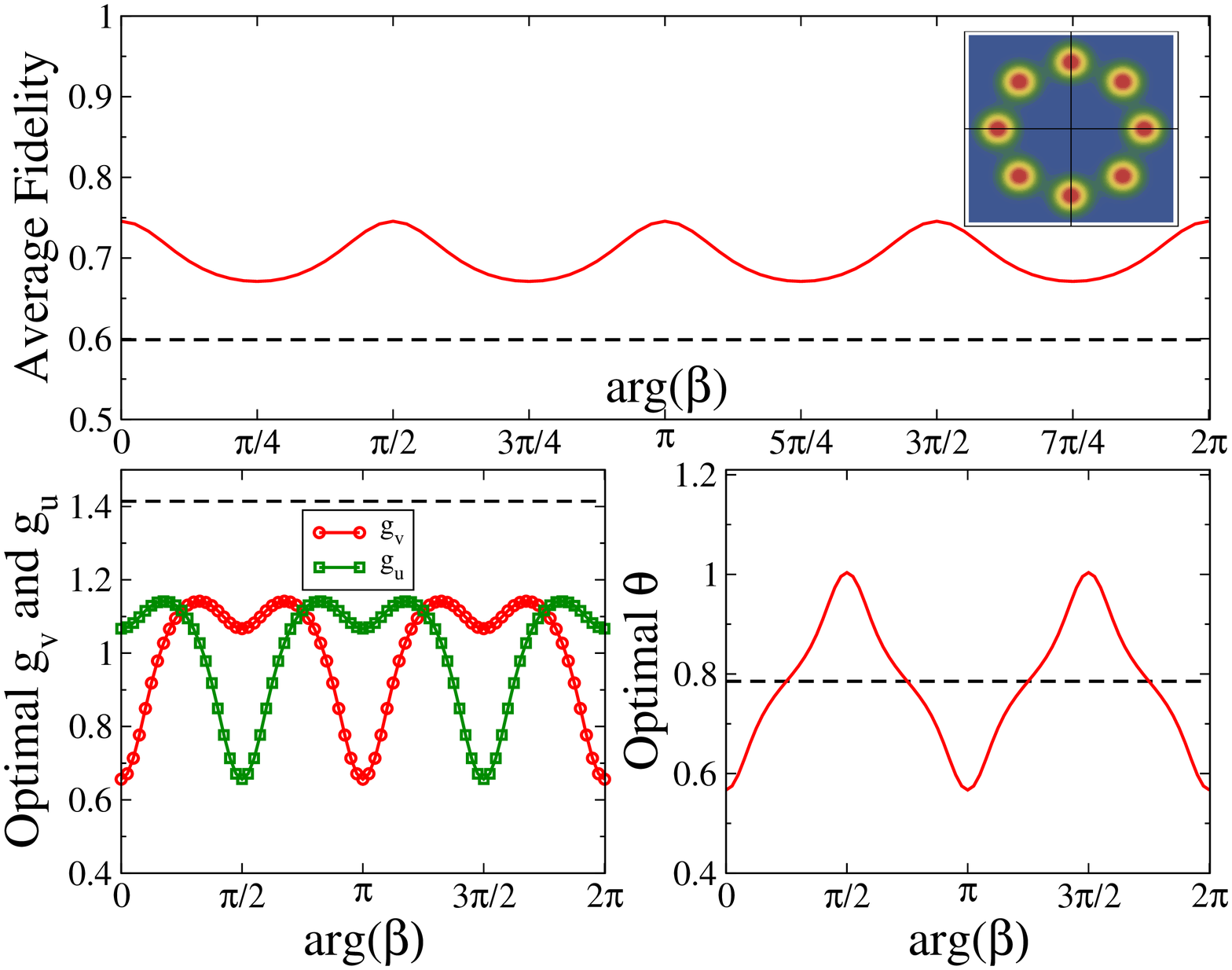}
\caption{\label{fig20}  We show a 
Gaussian distribution with variance $1/(2\lambda)$, $\lambda=2.0$, and 
mean $\beta = |\beta|e^{\arg(\beta)}$ with $|\beta|=1.5$. The
squeezing of the channel is $r=0.2$. Top panel: Optimal average
fidelity as a function of $\arg(\beta)$ (solid curve). 
The inset shows the Gaussians with greatest
(centered on the real and imaginary axis) 
and lowest ($\arg(\beta)=\pm 45^o$ and $\arg(\beta)=\pm 135^o$) optimal average fidelities.
Left lower panel: The optimal $g_v$ and $g_u$, the values of which are in quadrature ($90^o$ dephasing). 
Note that they are only equal for
$\arg(\beta)=\pm 45^o$ and $\arg(\beta)=\pm 135^o$, i.e., when $|Re(\beta)|=|Im(\beta)|$.
Right lower panel: The optimal $\theta$, which equals $\theta=45^o$ at the same points
where $g_v=g_u$. All dashed curves represent the corresponding values for the original CVTP.}
\end{center}
\end{figure}

If we work with Gaussian distributions displaced by $\beta\neq 0$, we cannot get a simple closed
solution to the optimization problem and we must rely on numerical methods. 
In Figs. \ref{fig19} and \ref{fig20} we show numerical computations giving the optimal average fidelity
and the optimal settings 
for Gaussians displaced in a variety of ways from the origin of the complex plane.

For the displaced Gaussians we observed that whenever 
$|Re[\beta]|=|Im[\beta]|$ the optimal parameters are such that $g_v=g_u$ and 
$\theta=\pi/4$, while $g_v\neq g_u$ and $\theta\neq\pi/4$
when $|Re[\beta]| \neq |Im[\beta]|$ (lower panels of Fig. \ref{fig20}). 
This result shows that for the great 
majority of the distributions here investigated the single parameter
optimization strategy is not enough to achieve the highest efficiency 
possible. Note also that for those cases Alice must employ an unbalanced
BS in order to achieve the greatest fidelity.

We have also noted that as we increase the squeezing $r$ and
the width of the distribution we approach the original CVTP fidelity
for distributions centered at any $\beta$. 
However, and quite surprising, if we displace the distribution 
along either the real or imaginary axis, the optimal average fidelity
has an asymptotic limit as we increase $|\beta|$ 
considerably better than the one predicted
for the original CVTP (left panels of Fig. \ref{fig19}). 
As we move the center of the distribution away 
from the real/imaginary axis, the asymptotic optimal average fidelity starts to approach the fidelity 
given by the original CVTP. For $\arg(\beta)=30^o$ we already have the asymptotic optimal average fidelity
indistinguishable from the original CVTP fidelity 
(right panels of Fig. \ref{fig19}).

\section{Discussions and Conclusion}

We have extensively studied how one can modify the original continuous variable teleportation
protocol \cite{Bra98} in order to increase its efficiency in teleporting coherent states 
by taking into account two facts inherently present in any actual implementation. 
The first one is the fact that Alice and Bob always deal with non-maximally 
entangled resources (finitely squeezed two-mode states) and the second one is related to the fact
that Alice's pool of possible coherent states to be teleported cannot cover the entire 
complex plane. 

After studying several different probability distributions for the pool of input states with Alice,
we showed that considerable gains in efficiency are achieved for all distributions 
if we introduce two slight modifications in the original setup. The first modification was the use of
a beam splitter whose transmittance could be changed at our will. This beam splitter was
employed to mix the state to be teleported with Alice's share of the entangled
resource, instead of the usual balanced beam splitter. The other modification was 
the possibility to freely choose the displacements in the quadratures, i.e., in position and momentum, of the 
output (teleported) state with Bob. By allowing these three actions to be independently adjusted once
the entanglement of the channel and the pool of input states are known, we were able to achieve 
considerable gains in efficiency when compared to that predicted by the original protocol.

We have also compared the three-parameter optimization strategy against the usual
one-parameter strategy, where the position and momentum
gains are not independently chosen  \cite{Bra01,Ide02,Bra99}. For certain types of distributions for the
pool of input states with Alice, namely, 
those centered at the vacuum state and with circular symmetry, we have shown that
the three-parameter strategy reduces to the one-parameter case. However, when the 
circular symmetry is broken, the three-parameter strategy is crucial in order to get a more 
efficient teleportation protocol. Indeed, we have shown that for 
circular symmetry broken distributions, the one-parameter strategy does not give significant
gains in efficiency when compared to the original teleportation protocol while 
the three-parameter strategy gives considerable gains.

In addition to important gains in efficiency with the three-parameter strategy,
we were also able to identify an interesting feature for distributions off-centered
from the origin of the complex plane but with the circular symmetry point lying on either the real or imaginary 
axis. We have shown that these distributions achieve the highest gain in performance
when compared to the equivalent distributions with symmetry points 
centered away from the real and imaginary axis.
Also, for distributions with circular symmetry points lying on the real and imaginary axis, 
we have shown that as we increase the distance of the circular symmetry point from the
origin, the optimal efficiency tends to a limiting value that is greater than the 
efficiency of the original teleportation protocol. This effect is more expressive for channels with
a low degree of entanglement and is absent for distributions with the symmetry point not
belonging to the real and imaginary axis.  We believe these interesting properties might be useful in 
the implementation of continuous variable quantum key distribution schemes based on coherent states
\cite{Ral99,Gra02,Leu02,Gra03,Nam03,Nam03b,Nam04,Nam06,Leu09,Leu10,Jou13}, 
where instead of transmitting the coherent state between the parties involved in
the key distribution scheme, with the transmission process adding noise and degrading
the signal, one teleports it using the optimal strategy here presented.

Furthermore, the calculations we made assumed a two-mode vacuum squeezed state
as our entanglement resource and a pool of input states given by coherent 
states. These choices were dictated by the fact that the usual 
resources employed in actual implementations of continuous variable
teleportation are described by these states \cite{Bra98,Fur98}. 
However, the formalism presented in Sec. \ref{secformalism} is quite 
general and can be easily adapted to any input state and any type of 
quantum channel. These changes are mathematically implemented by simply
substituting the input and entangled states' expansion coefficients 
in the position basis, equations~(\ref{input}), (\ref{canal}), (\ref{in}), and
(\ref{ch}), with the corresponding ones for the new states.  

Finally, we would like to point out to a particular extension of 
the research here presented that might prove fruitful. It is an extensive
analysis of the three parameter optimization strategy for several pool of input
states assuming that Alice and Bob share 
quantum channels given by \textit{mixed} states. Since decoherence, noise, and attenuation drive
pure entangled states to mixed ones after a sufficient exposure time,
it would be interesting to investigate whether the same techniques here presented
can be helpful in improving the efficiency of a continuous variable teleportation
protocol that employs non pure channels.

\section*{Acknowledgments}
FSL and GR thank CNPq (Brazilian National Science Foundation) for
funding this research. GR also thanks 
FAPESP (State of S\~ao Paulo Science Foundation) 
and CNPq/FAPESP for financial support through the 
National Institute of Science and Technology for Quantum Information (INCT-IQ).


\begin{thebibliography}{99}

\bibitem{Ben93}  C. H. Bennett \textit{et al.}, Phys. Rev. Lett. 70 (1993) 1895.

\bibitem{Vai94} L. Vaidman, Phys. Rev. A 49 (1994) 1473.

\bibitem{Bra98} S. L. Braunstein and H. J. Kimble, Phys. Rev. Lett. 80 (1998) 869.

\bibitem{Ral98} T. C. Ralph and P. K. Lam, Phys. Rev. Lett. 81 (1998) 5668.

\bibitem{Bow97} D. Bouwmeester \textit{et al.}, Nature (London) 390 (1997) 575.

\bibitem{Bos98} D. Boschi \textit{et al.}, Phys. Rev. Lett. 80 (1998) 1121. 

\bibitem{Fur98} A. Furusawa \textit{et al.}, Science 282 (1998) 706.

\bibitem{Bow03a} W. P. Bowen \textit{et al.}, Phys. Rev. A 67 (2003) 032302.

\bibitem{Kim03} T. C. Zhang \textit{et al.}, Phys. Rev. A 67 (2003) 033802.

\bibitem{Bra05} S. L. Braunstein and P. van Loock,
Rev. Mod. Phys. 77 (2005) 513.

\bibitem{Bra00} S. L. Braunstein, C. A. Fuchs, and H. J. Kimble, J. Mod. Opt. 47 (2000) 267. 

\bibitem{Bar98} H. Barnum, (PhD Thesis, University of New Mexico, Albuquerque, NM, USA, 1998).

\bibitem{Bal98}  L. E. Ballentine, Quantum Mechanics: A Modern Development 
(World Scientific, Singapore, 1998).

\bibitem{Bra01} S. L. Braunstein, Ch. A. Fuchs, H. J. Kimble, and P. van Loock, Phys. Rev. A 64 (2001) 022321.

\bibitem{Ide02} T. Ide, H. F. Hofmann, A. Furusawa, and T. Kobayashi, Phys. Rev. A 65 (2002) 062303.

\bibitem{Mis10} L. Mista Jr., R. Filip, and A. Furusawa, Phys. Rev. A 82 (2010) 012322.

\bibitem{Bow03} W. P. Bowen \textit{et al.}, IEEE J. Sel. Top. Quantum Electron. 9 (2003) 1519.

\bibitem{Pol99} R. E. S. Polkinghorne and T. C. Ralph, Phys. Rev. Lett. 83 (1999) 2095.

\bibitem{Bra99} P. van Loock and S. L. Braunstein, Phys. Rev. A 61 (1999) 010302(R).

\bibitem{Ral99} T. C. Ralph, Phys. Rev. A 61 (1999) 010303(R).

\bibitem{Gra02} F. Grosshans and Ph. Grangier, Phys. Rev. Lett. 88 (2002) 057902. 

\bibitem{Leu02} Ch. Silberhorn, T. C. Ralph, N. L\"utkenhaus, and G. Leuchs, Phys. Rev. Lett. 89 (2002) 167901.


\bibitem{Gra03} F. Grosshans \textit{et al.}, Nature (London) 421 (2003) 238.

\bibitem{Nam03} T. Hirano \textit{et al.} Phys. Rev. A 68 (2003) 042331.

\bibitem{Nam03b} R. Namiki and T. Hirano, Phys. Rev. A 67, 022308 (2003).

\bibitem{Nam04} R. Namiki and T. Hirano, Phys. Rev. Lett. 92 (2004) 117901.

\bibitem{Nam06} R. Namiki and T. Hirano, Phys. Rev. A 74 (2006) 032302.

\bibitem{Leu09} D. Elser \textit{et al.}, New J. Phys. 11 (2009) 045014.

\bibitem{Leu10} D. Sych and G. Leuchs, New J. Phys. 12 (2010) 053019.

\bibitem{Jou13} P. Jouguet \textit{et al.}, Nature Photon. 7 (2013) 378.


\end{thebibliography}
\end{document}